\documentclass[journal=apchd5,manuscript=article]{achemso}

\usepackage{chemformula} 
\usepackage[T1]{fontenc} 
\usepackage{siunitx}
\usepackage{filecontents}
\usepackage{xr}

\author{Daniel B\"{o}ning}
\affiliation{Max Planck Institute for the Science of Light, 91058 Erlangen, Germany}
\altaffiliation{these authors contributed equally}
\author{Franz-Ferdinand Wieser}
\affiliation{Max Planck Institute for the Science of Light, 91058 Erlangen, Germany}
\alsoaffiliation
{Friedrich Alexander University Erlangen-Nuremberg, 91054 Erlangen, Germany}
\altaffiliation{these authors contributed equally}
\author{Vahid Sandoghdar}
\affiliation{Max Planck Institute for the Science of Light, 91058 Erlangen, Germany}
\email{vahid.sandoghdar@mpl.mpg.de}
\affiliation
{Max Planck Institute for the Science of Light, 91058 Erlangen, Germany}
\alsoaffiliation
{Friedrich Alexander University Erlangen-Nuremberg, 91054 Erlangen, Germany}

\title{Polarization-encoded co-localization microscopy at cryogenic temperatures}

\keywords{single-molecule localization microscopy, cryogenic light microscopy, polarization, DNA origami}

\begin{document}

\begin{abstract}
Super-resolution localization microscopy is based on determining the positions of individual fluorescent markers in a sample. The major challenge in reaching an ever higher localization precision lies in the limited number of collected photons from single emitters. To tackle this issue, it has been shown that one can exploit the increased photostability at low temperatures, reaching localization precisions in the sub-nanometer range. Another crucial ingredient of single-molecule super-resolution imaging is the ability to activate individual emitter within a diffraction-limited spot. Here, we report on photoblinking behavior of organic dyes at low temperature and elaborate on the limitations of this ubiquitous phenomenon for selecting single molecules. We then show that recording the emission polarization not only provides access to the molecular orientation, but it also facilitates the assignment of photons to individual blinking molecules. Furthermore, we employ periodical modulation of the excitation polarization as a robust method to effectively switch fluorophores. We bench mark each approach by resolving two emitters on different DNA origami structures. 
\end{abstract}


\section*{Introduction}

With the advent of super-resolution methods, optical microscopy has provided fascinating new insights into the sub-cellular domain and has become an indispensable tool in elucidating the structure and function of biological systems at the nanoscale\cite{Weisenburger2015}. The high specificity and spatial resolution of fluorescence imaging has the potential to deliver further information on the molecular architecture of proteins and their complexes even in a native environment, e.g. membrane proteins or protein aggregates implicated in diseases. Recently, it has been recognized that super-resolution microscopy performed at cryogenic temperatures can be of great value \cite{Weisenburger2017,LeGros2009,Kaufmann2014,Li2015,Nahmani2017,Wang2019,Furubayashi2019,Chang2014,Dahlberg2020,Hoffman2020}. The main advantage of this approach stems from the fact that photochemistry is considerably slowed down at low temperatures. As a result, each fluorophore can emit more than two orders of magnitude more photons than at room temperature before it photobleaches. This translates into a higher localization precision and, thus, better resolution in co-localization of several fluorophores. Another important benefit of cryogenic light microscopy is its potential for combination with cryogenic electron microscopy and correlative microscopy\cite{Chang2014,Tuijtel2019,Moser2019,Dahlberg2020,Hoffman2020}. While cryogenic super-resolution microscopy in organic crystals predates conventional super-resolution microscopy by about a decade \cite{Guettler1994, Hettich2002}, its use in biologically relevant applications has been a theme of research only recently\cite{Kaufmann2014,Li2015,Weisenburger2017,Nahmani2017,Wang2019}. 

The best resolution in biological super-resolution microscopy has been reported by Cryogenic Optical Localization in three Dimensions (COLD), reaching Angstrom resolution of up to four fluorophores on a single protein\cite{Weisenburger2017}. In that work, cases of exceptionally slow blinking were used to identify brightness levels of the individual emitters and their combinations on a single protein. However, this strategy limits the yield of the experimental procedure because as we discuss in this work, most molecules show faster photophysics. To understand and tame this difficulty, we have performed more detailed photophysics studies at liquid helium temperature. Furthermore, we have exploited the polarization degree of freedom associated with the dipole moments of the fluorophores as a resource for separating their signals and discuss its influence on localization accuracy\cite{Enderlein2006,Engelhardt2011}.

\section*{Blinking of red fluorescent organic dyes at cryogenic temperatures}
Naturally occurring stochastic blinking of fluorophores is a ubiquitous phenomenon with different physical origins, e.g. intersystem crossing to triplet states, charge trapping, conformational changes or transient binding\cite{Dickson1997,Zondervan2003,Clifford2007,Ha2012}. Blinking offers a convenient universal scheme for nanoscopic studies that involve a handful of molecules within a range of a few nanometers, but the different time scales, spanning from microseconds to minutes, can make the distinction of a large number of emitters in a diffraction-limited spot extremely difficult. 

In the simplest approach, one records videos from a field-of-view of about $\SI{1000}{\mu m^2}$ and examines the time trace from each diffraction-limited spot\cite{Weisenburger2017}. In the ideal case one obtains $2^N$ intensity levels corresponding to $N$ active fluorophores such that it is possible to find frames where only one fluorophore remains on, and can thus be localized. By repeating this procedure for videos as long as tens of minutes or hours, one gathers sufficient data to reach sub-nanometer localization precision for each fluorophore and, hence, resolve their relative positions. For this procedure to work, it is important to know about the switching rates of the fluorophore under the specific experimental conditions.

\begin{figure}[ht]
\centering
\includegraphics[width=\textwidth]{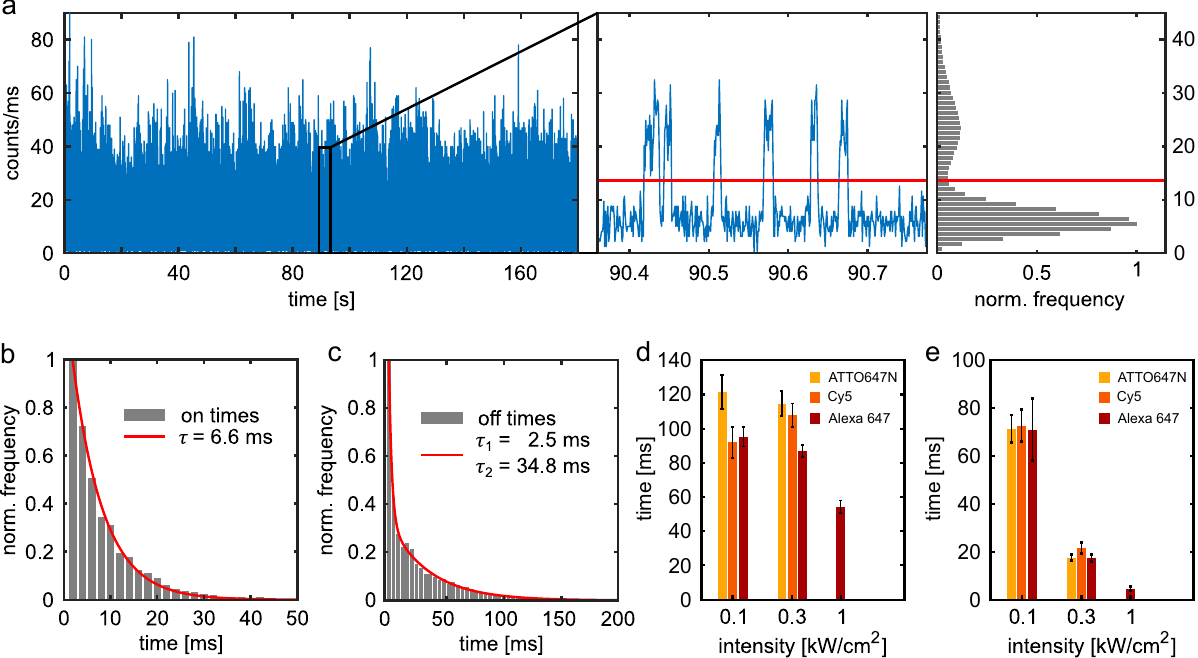}
\caption{\label{Figure1}
Photoblinking at cryogenic temperatures. a) Example time trace of a single Alexa Fluor 647 molecule in poly-vinyl alcohol binned in $\SI{1}{ms}$ intervals. A zoom-in shows short emission bursts followed by longer dark periods. The red line indicates the threshold used to compute on- and off-times. The right plot shows the intensity histogram of the signal bursts. b) Histogram of the on-times of the trace shown in a) with an exponential fit. c) Histogram of off-times for the same trace. Here, the statistics is best described by a bi-exponential function with a short and a long component. d) Summary of long off-times for the different dye species at different excitation intensities. e) Summary of the corresponding on-times.}
\end{figure}

While the most commonly used organic dyes have been well characterized at room-temperature\cite{Dempsey2011}, there is still little information on the photophysics of fluorescent labels at low temperatures. A quantitative understanding of this topic requires a thorough study of many parameters regarding the fluorophore and its environment and is beyond the scope of our work. Nevertheless, we attempt to present a flavor of the phenomena at hand for the three common red-fluorescent dyes Alexa Fluor 647, Cy5 and ATTO647N in a poly-vinyl alcohol (PVA) host matrix. We perform single-photon counting to obtain the time traces of single molecules and extract the duration of on- and off-periods. As shown in Figure\,\ref{Figure1}a, application of a threshold at 3 standard deviations from the mean background photon count of the brightness histogram allows us to flag an event as on or off. We verified that small variations of the threshold did not change the obtained on- and off-times significantly and also found good agreement with time constants computed from the analysis of the auto-correlation function. 

In Figure\,\ref{Figure1}b,c we find that in the case of Alexa Fluor 647 the duration of on-times follows an exponential distribution whereas off-times are best described by a bi-exponential function with a short and a long time constant. Figure\,\ref{Figure1}d shows the (long) off-times for three different excitation intensities, revealing little dependence over the investigated range. The on-times, however, decrease down to $\SI{5}{ms}$ at elevated excitation intensities (exceeding $\SI{1}{kW/cm^2}$) as shown in Figure\,\ref{Figure1}e, a phenomenon which is exploited in a typical (d)STORM situation and can even be chemically engineered\cite{Steinhauer2008}. As a rule of thumb, the larger the off-on ratio, the higher the probability to localize individual emitters from a set of many within a diffraction-limited spot. However, considerations such as the relation between the integration time and the on- and off-times should also be taken into account\cite{Pennacchietti2017}. 

The three dyes investigated here show very similar transition rates and brightnesses. We point out that the characteristic exponential blinking kinetics are sometimes interrupted by long emission bursts or long dark periods in all cases, possibly indicating reversible changes of the triplet state lifetime\cite{Veerman1999} or the molecular configuration\cite{Weston1999}. This behavior is in line with a similar conclusion found for ATTO647N at room temperature, where the blinking statistics were shown to depend on the environment\cite{Zondervan2003,Clifford2007} with primary sources of charge transfer, triplet states and radical ion states.

\begin{figure}[ht]
\centering
\includegraphics[width=0.5\textwidth]{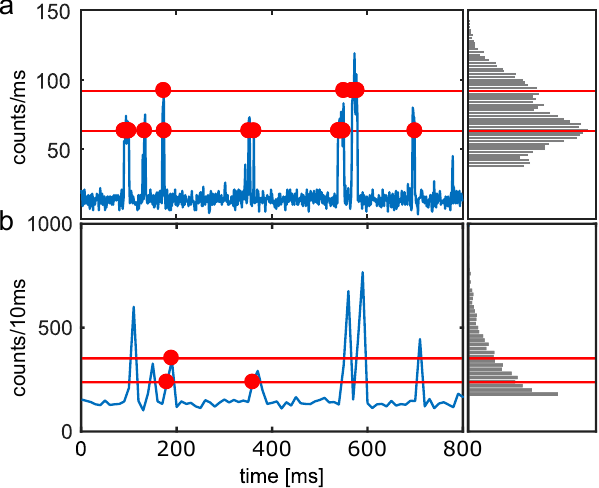}
\caption{\label{Figure2}
Photoblinking of a nanoruler containing two Alexa Fluor 647 dyes. a) A short interval of the time trace sampled at $\SI{1}{ms}$. Individual blinking events are clearly visible and can be assigned to a fluorophore based on the brightness. Red lines indicate the levels found by the algorithm used in Ref.\citenum{Weisenburger2017}, and dots indicate data points within the respective shot-noise bounds. b) The same trace sampled at $\SI{10}{ms}$. Individual blinking events are still resolved but assignment to single molecules fails as the histogram starts to show an exponentially decaying continuum of brightness values. }
\end{figure}

Next, we briefly present the blinking behaviors of two Alexa Fluor 647 dyes placed on an origami nanoruler. As displayed in Figure\,\ref{Figure2}a, one can clearly identify two brightness levels at a temporal resolution of $\SI{1}{ms}$ where the blinking events are sufficiently oversampled. The brightness histogram of the same trace shows a continuum without any distinct levels if binned to $\SI{10}{ms}$, even though the blinking of individual fluorophores is still temporally resolved. This study emphasizes that frame acquisition slightly faster than the on-time is required for assigning an intensity level to a given molecule, limiting the performance of this method. 

\section*{Selection via the emission polarization}

To alleviate the difficulty of identifying two fluorophores based on their fluctuating brightness levels, we now exploit the polarization degree of freedom \cite{Gould2008,Cruz2016}. The emission dipole moment of a dye molecule is usually well defined with respect to its backbone such that the polarization of the radiated field can directly report on its orientation. While in room-temperature aqueous environments the fluorescent label is free to rotate about its linker, the orientation of an emitter is typically fixed at cryogenic temperatures. Hence, the emission dipole orientation in the image plane can be determined from a measurement of the emission intensities $I_{x,y}$ projected along two orthogonal lateral axes $x$ and $y$ according to
\begin{equation}
\theta = \arctan\sqrt{\frac{I_x}{I_y}},\qquad \theta\in\left[0, \pi/2\right]\,.
\label{Equation1}
\end{equation}
Given that the orientations of several individual molecules are independent, one can distinguish their signals if their polarization angles are sufficiently spaced. 

\begin{figure}[ht!]
\centering
\includegraphics[width=0.5\textwidth]{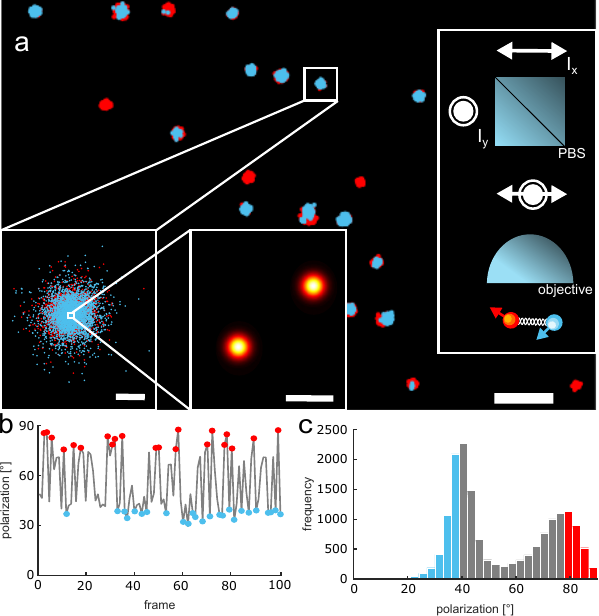}
\caption{\label{Figure3} Polarization-enhanced localization analysis. a) Cryogenic super-resolution image of DNA nanorulers labeled with two ATTO647N dyes. The localizations of each fluorophore are assigned according to the polarization and shown as red and blue dots, respectively. The zoomed-in view shows a single nanoruler with its localizations and the highly resolved image obtained by taking their average. The inset shows a sketch of the polarization-resolved detection. Scale bars: $\SI{3}{\mu m}$, $\SI{100}{nm}$ and $\SI{3}{nm}$. b) Polarization of the nanoruler shown in a) with red and blue dots indicating frames retained for localization. c) Histogram plot of the measured polarization with two clearly distinguishable peaks.}
\end{figure}

To implement this idea, we separated the two polarizations along the $x$ and $y$ directions with a polarizing beam splitter and directed them to two separate synchronized cameras (see inset in Figure\,\ref{Figure3}a). Figure\,\ref{Figure3}b shows an exemplary time trace of the polarizations extracted by applying Eq.\,\ref{Equation1} to signals $I_{x,y}$. Although both label molecules might be blinking during one frame, the extreme signals marked by the blue and red circles clearly point to situations where only one molecule was on. We only use the data from these frames for localization to avoid any overlap. Figure\,\ref{Figure3}c presents the same data as a histogram. We see that in contrast to the brightness histogram (see Figure\,\ref{Figure2}b), the dipole angle follows a symmetrical distribution that clearly identifies two distinct polarizations, greatly facilitating the assignment of the signals to individual molecules, indicated by the red and blue portions. The correspondingly color-coded spots in Figure\,\ref{Figure3}a display the registered localizations from the two fluorophores of the nanorulers. A zoom into one of the spots shows a strong overlap of the localized positions, which after averaging yields two spots separated by $\SI{7.5}{nm}$. In this example, the two point-spread functions (PSFs) could be clearly separated because the dipole orientations of the two molecules on a DNA origami were sufficiently different. 

\begin{figure}[ht]
\centering
\includegraphics[width=0.5\textwidth]{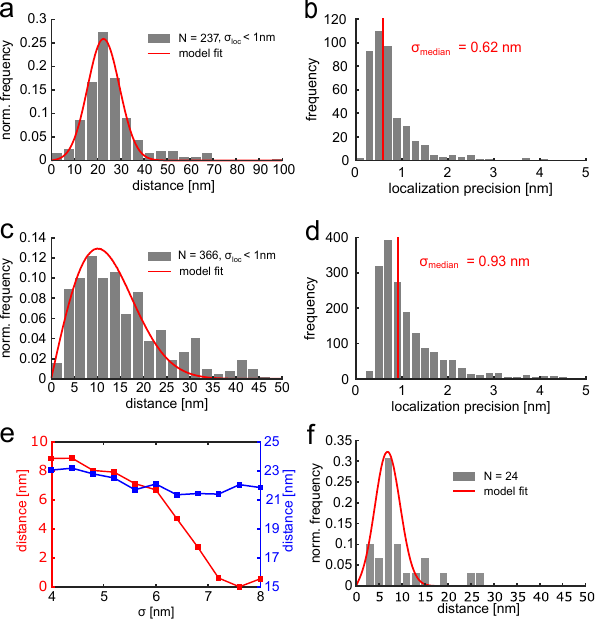}
\caption{\label{Figure4} Nanometer distance measurement on DNA origamis. a) Distribution of distances for the $\SI{22.8}{nm}$ nanoruler and its model fit. b) Localization precision of the $\SI{22.8}{nm}$ rulers in a). c) Distribution of distances for the $\SI{6.5}{nm}$ nanoruler and its model fit.  d) Localization precision for the data set in c). e) Outcome of fitted distances for a) (blue) and c) (red) as a function of the model parameter $\sigma$. f) Filtering the histogram in c) for the brightest fluorophores and small relative dipole orientations leads to a significant narrowing of the distribution.}
\end{figure}

To state a statistically meaningful distance between the labeling sites of a nanostructure, we examine a large number of particles. Figure\,\ref{Figure4}a,c display the distribution of the separations deduced from measurements on origami samples where two Alexa Fluor 647 molecules were placed at a design separation of $\SI{22.8}{nm}$ (Tilibit) and two ATTO647N molecules were placed at a design separation of $\SI{6.5}{nm}$ (GATTAQuant), respectively. We now discuss the various effects that determine the shapes of these distributions.

We point out that our samples have a linear architecture so that the two-dimensional projection of two PSFs should always report distances equal to or smaller than the design distance. Furthermore, the PSF of a molecule depends on the orientation of its dipole moment\cite{Enderlein2006,Engelhardt2011}. Unless the dipole fully lies in the lateral or the axial planes, its PSF is skewed leading to systematic errors in localization and thus an apparent shift of the molecular center of mass (see Figure S2). Thus, to account for the distribution of the occurrence frequencies in the histograms, one has to consider the localization uncertainty, which can be estimated as the quadratic sum of independent contributions. These include a statistical localization error $\sigma_{\rm loc}$ shown in Figure\,\ref{Figure4}b,d, an image registration error $\sigma_{\rm reg}$ (see Figure S3), a residual sample drift $\sigma_{\rm dft}$ and an average error $\sigma_{\rm dip}$ due to the fixed dipole orientation (see Figure S2), yielding
\begin{equation}
\sigma^2 = 2\cdot(\sigma_{\rm loc}^2+\sigma_{\rm dip}^2)+\sigma_{\rm reg}^2+\sigma_{\rm dft}^2\,.
\end{equation}
Considering the estimated error, we can now fit the localization distributions using a bivariate normal distribution with non-zero mean, also called Rician distribution\cite{Churchman2006,Niekamp2019}. In this model, if the separation $d$ between the two molecules is much larger than the localization uncertainty $\sigma$, the distribution closely resembles a pure Gaussian so that the expectation value and the median coincide. However, if $d\approx\sigma$ the distribution acquires an asymmetric tail and can no longer be described by a simple median. The red curve in Figure\,\ref{Figure4}a results from the application of this model to our measurements for the nanorulers with dyes separated by $\SI{22.8}{nm}$, yielding $d=21.4\pm 2.0\,\textnormal{nm}$, in agreement with the outcome of a Gaussian fit estimate $22.8\pm 0.4\,\textnormal{nm}$ and the median $23.7\pm 0.8~\textnormal{nm}$. Here, 414 of the 2854 localized PSFs showed a clear polarization signature, whereby 237 were sufficiently well separated and well localized. This corresponds to a yield of $8\%$ of all detections, excluding ones with only one fluorophore. We remark that some of the observed fluorescence spots stem from impurities spread on the sample and are used for image registration. For the nanoruler sample with dyes separated by $\SI{6.5}{nm}$, the model estimates $d=6.6\pm 2.3~\textnormal{nm}$, whereas the median lies at $14.0\pm 0.8~\textnormal{nm}$. In Figure\,\ref{Figure4}e, we plot the dependence of the extracted distance from data fits on the input value of $\sigma$, verifying that the distance assignment is very sensitive to $\sigma$ for the smaller nanoruler\cite{Niekamp2019}. 

The shape of the distance distributions using an extended model accounting for the 3D orientation of origamis (see Figure S4) indicates that the origami structures mostly lie parallel to the surface. Nevertheless, the dipole moments of the individual molecules could be arbitrarily oriented in space. The radiation of an axial dipole moment placed at a dielectric interface is emitted into larger angles, leading to a doughnut-shaped PSF\cite{Enderlein2006}. It follows that the fluorescence of such a molecule is less efficiently excited and collected by an air objective and the molecule appears less bright. A mixture of in-plane and out-of-plane dipole components leads to an asymmetric PSF, thus, introducing a systematic localization error if the PSF is simply fitted by a Gaussian function. However, the PSFs of two dipoles with the same orientations are shifted synchronously, leaving their center-to-center separation almost unaffected even if the fit function is not ideally adapted. Figure\,\ref{Figure4}f shows that, indeed, by selecting bright dipoles and small relative angles between the two fluorophores of the data presented in Figure\,\ref{Figure4}b, we arrive at narrower and more symmetric distributions. We remark that the complication caused by the 3D orientation of the dipole moment could be addressed more rigorously by direct measurement of the complete orientation\cite{Furubayashi2019} or by filtering the azimuthal contributions of the PSF with a phase mask\cite{Backlund2016}. 

We also remark that the success of polarization selection comes at the cost of a lower signal in each channel since we have to split the emission from single molecules. To maintain a good signal-to-noise ratio, we placed a mirror at the substrate surface in order to also capture the light that is emitted away from the microscope objective\cite{Moal2007,Heil2018}. Besides enhancing the excitation and collection efficiencies (see Figure S1), the mirror also eliminates autofluorescence of the glass substrates, which would introduce a considerable background. However, it also affects the PSF and therefore the distance measurement (see Figure S2). 

\section*{Switching molecules via the excitation polarization}

\begin{figure}
\centering
\includegraphics[width=0.8\textwidth]{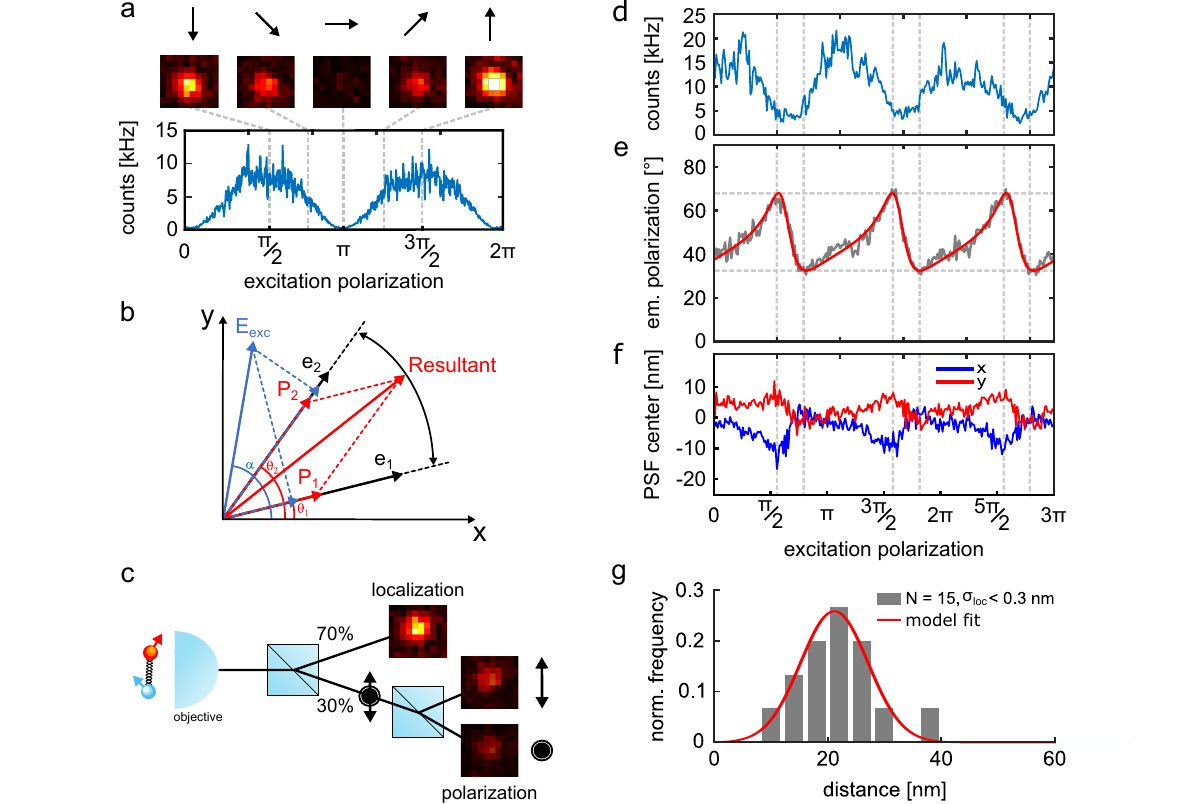}
\caption{\label{Figure5}Schematics of polarization modulation microscopy. a) Modulation of a single molecule fluorescence by rotating the excitation polarization, indicated by the arrows. The brightness shows a $\cos^2\alpha$ behavior. b) Diagram illustrating the polarization of the total signal from two fluorophores ($P_{1,2}$) with dipole orientations $\textbf{e}_1$ and $\textbf{e}_2$ excited by light of linear polarization along $\textbf{E}_{\rm exc}$. c) Sketch of the optical setup. d) The detected brightness of a single nanoruler shows oscillations as the input polarization is rotated. e) The polarization trace of the total emission shows correlated oscillations and clearly reveals the presence of two fluorophores at $32.5^\circ$ and $68.1^\circ$. The red line represents a model fit. f) We track the center of the PSF and observe an oscillation between the positions of the fluorophores that is also correlated with the brightness and the polarization trace. g) Distances measured on the $\SI{22.8}{nm}$ nanoruler.}
\end{figure}

Selection of individual fluorophores via a stochastic phenomenon such as blinking is intriguing and broadly applicable. However, deterministic activation by an external command provides more control. We now apply a convenient and general approach based on the modulation of the excitation polarization with respect to the absorption dipole moments of the fluorophores\cite{Hafi2014,Backer2016,Hulleman2018}. The excitation probability of a linear dipole moment aligned at an angle $\theta$ follows a $\cos^2(\theta-\alpha)$ dependence, where $\alpha$ denotes the polarization angle of the incoming linearly polarized light. Thus, a single molecule can be completely turned off if $\alpha=\theta \pm 90^\circ$. Figure\,\ref{Figure5}a displays an example of the sinusoidal dependence of the detected fluorescence from a single molecule on the incident polarization angle.

We now excite a nanoruler carrying two dye molecules with absorption dipole moments along $\theta_1$ and $\theta_2$, respectively (see Figure\,\ref{Figure5}b). As the excitation polarization angle $\alpha$ is rotated, the total brightness of the detected light can be expressed as a superposition of the components $P_{1,2}$ originating from the two fluorophores along the unit vectors $\textbf{e}_1$ and $\textbf{e}_2$ which signify the directions of their dipole moments, respectively. We note that although the absorption and emission dipoles of organic dye are generally not aligned\cite{Lakowicz2006}, we assume this to be the case for the sake of simplicity here.

Such a linear decomposition is valid as long as the molecules are not saturated and their response to the excitation intensity is linear. Hence, by rotating the polarization of the excitation light, we expect the detected signal to show a periodic modulation. Here, the resultant vector associated with the total emission will be confined between the extreme cases of angles $\theta_1$ and $\theta_2$, where only one molecule contributes to the signal because the excitation happens to be perpendicular to the other one.

To realize such measurements, we inserted a rotatable linear polarizer in the excitation beam path to generate linearly polarized light of arbitrary orientation in the sample x-y plane. For detection, the light originating from the sample was split with a non-polarizing 30:70 beam-splitter, whereby $30\%$ of the light was further split with a Wollaston prism onto two regions of a camera to analyze the emission polarization. The remaining light was projected to a second synchronized camera to perform localization (see Figure\,\ref{Figure5}c). With this imaging scheme, localization and emission polarization can be measured independently while each individual localization can be assigned to a polarization state. An important advantage of this approach is that it is largely independent of the blinking dynamics. The excitation polarization can be rotated very slowly as long as the fluorophores do not photobleach during one $180^\circ$ rotation. This allows for long integration times and an increased signal-to-noise ratio per camera frame. Another advantage of the scheme is that localization can be performed without the need for image registration.

In Figure\,\ref{Figure5}d-g, we present an example of a nanoruler carrying two dyes. Here, each frame was recorded for 3s, resulting in a single-molecule localization precision of a few nanometers per frame. Three frames were then averaged for each excitation polarization angle, which was incremented in steps of $2^\circ$. Figure\,\ref{Figure5}d shows the total brightness recorded from the two molecules on a nanoruler. While a periodic modulation is evident, the angles $\theta_1$ and $\theta_2$ are not easily identifiable. However, if we exploit the information about the emission polarization, i.e. our knowledge of the vectors $\textbf{e}_1$ and $\textbf{e}_2$, we can assign a polarization to the detected fluorescence as illustrated in Fig.\,\ref{Figure5}b. Figure\,\ref{Figure5}e shows that, indeed, the angle attributed to the total emission is confined between two extrema. In the special case that the two absorption dipole moments are perpendicular to each other and lie in the substrate plane, the vertical axis in Figure\,\ref{Figure5}e would cover the full range of $0-90^\circ$. We note that comparing Figure\,\ref{Figure5}d with Figure\,\ref{Figure5}e, we also find a clear correlation between brightness and emission polarization. Having identified the conditions where only one fluorophore is on, we can now localize it on the second camera. 


The camera images of the cases where both molecules contribute also remain useful because they help obtain a robust fit to the outcome of a fit according to the vectorial model illustrated in Figure\,\ref{Figure5}b. In particular, we expect the PSF of such intermediary states to wander between two extreme positions. Indeed, the x- and y-displacements of the recorded PSF shown in Figure\,\ref{Figure5}f reveals that the center-of-mass of the fluorescence spot moves back and forth between two locations as the excitation polarization rotates. By analyzing these data, we could determine the distance between the two fluorophores to be $21.1\pm 2.0~\textnormal{nm}$, in agreement with the previous result (see Figure\,\ref{Figure5}g).

The controlled switching of single molecules alleviates the experimental work because we can use long camera integration times and work with much lower light levels as required for stochastic switching. Furthermore, polarization measurement and localization are now separated, eliminating the errors associated with image registration between two cameras. The only requirement for this method is linearly polarized absorption and emission dipole moments of the fluorescent labels, regardless of their other photophysical properties such as blinking. 

\section*{Conclusions}

Cryogenic optical localization has reached nanometer resolution making it a valuable tool for structural biology and other applications in physics and material science, e.g. localizing defects and color centers in 2D materials. We have shown that including the polarization degree of freedom in the localization analysis allows for more robust assignment of fluorescence photons to individual emitters. Aside from boosting the localization accuracy, this approach also provides direct access to molecular orientations which can be useful in the context of studying agglomeration or oligomerization of proteins\cite{Tyakko2010,Kampmann2011,Camacho2018,Ding2020}. Furthermore, we have shown that polarization can be used to achieve controllable switching, which is largely independent of the stochastic blinking, works with lower light levels and allows for longer camera integration times. The implementations of the ideas in our work are straightforward and not restricted to specific photophycial properties. We have shown the potential of this technique for imaging nanostructures containing two molecules. Future efforts will tackle problems with many fluorophores, where the current work could also be combined with SOFI\cite{Moser2019} or sparsity-enhancing algorithms\cite{Hafi2014}.

\begin{acknowledgement}
We are grateful to the Max Planck Society for financial support. We acknowledge Tobias Utikal for assistance with cryogenic experiments and Alexander Gumann for preparing mirror-enhanced substrates.
\end{acknowledgement}

\begin{suppinfo}

The following files are available free of charge.
\begin{itemize}
  \item Supplementary information. Detailed description of the optical setup, sample preparation, data acquisition and image analysis. Extended model including 3D orientation of particles and estimation of distance uncertainty. Theoretical calculation of mirror enhancement and localization uncertainty due to dipole orientation
\end{itemize}

\end{suppinfo}

\bibliography{references}

\end{document}


\tableofcontents

\section*{Cryogenic microscope}

All measurements were performed in a cryogenic microscope where the sample is kept at liquid helium temperature. Fluorescence is excited in wide-field configuration with circularly polarized light at $\SI{640}{nm}$ from a diode laser (iBeamSmart, Toptica) and collected through a long-working distance objective with a numerical aperture of 0.9 (MPLAN 100x, Mitutoyo) mounted in vacuum. The laser light is filtered with a  $\SI{647}{nm}$ dichroic mirror (RazorEdge, Semrock) and further suppressed with $\SI{650}{nm}$ long-pass filters (Thorlabs). In order to determine the switching rates of the stochastic on-off blinking we use single-photon counting with a $30:70$ beam-splitter dividing the signal between an EMCCD camera (iXon, Andor) and an avalanche photo diode (Lasercomponents). 
For super-resolution imaging a polarizing beam-splitter (Thorlabs) is used together with a second EMCCD camera and acquisition of frames is synchronized via common trigger pulses.
Polarization modulation nanoscopy was performed in a slightly modified optical setup. The unmodified point-spread function is extracted from one camera receiving $70\%$ of all photons. The weights in each frame are determined from the remaining $30\%$ of photons projected into a polarization-resolved channel on two halves of a second camera chip using a Wollaston prism.
We use custom designed dielectric mirrors (Laseroptik) and low angles of incidence on other components to minimize polarization-dependent phase shifts
throughout the imaging optics. This turned out to be sufficient for our co-localization analysis. More quantitative studies of molecular orientation
may require to take the distortion of polarization in high-NA collection into account\cite{Gould2008,Cruz2016}.

\section*{Mirror enhancement}
\begin{figure}[ht]
\centering
\includegraphics[width=0.5\textwidth]{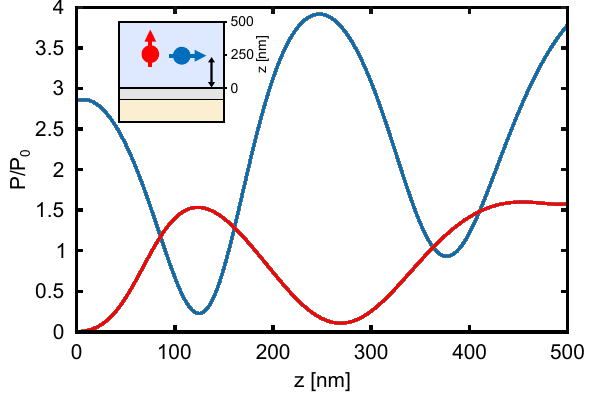}
\caption{\label{SuppFig1} Mirror enhancement of collection efficiency.  The graph shows the power radiated into the acceptance angle of the objective for horizontal (blue) and vertical (red) dipoles normalized to an interface without gold layer.}
\end{figure}
\begin{figure}[H]
\centering
\includegraphics[width=\textwidth]{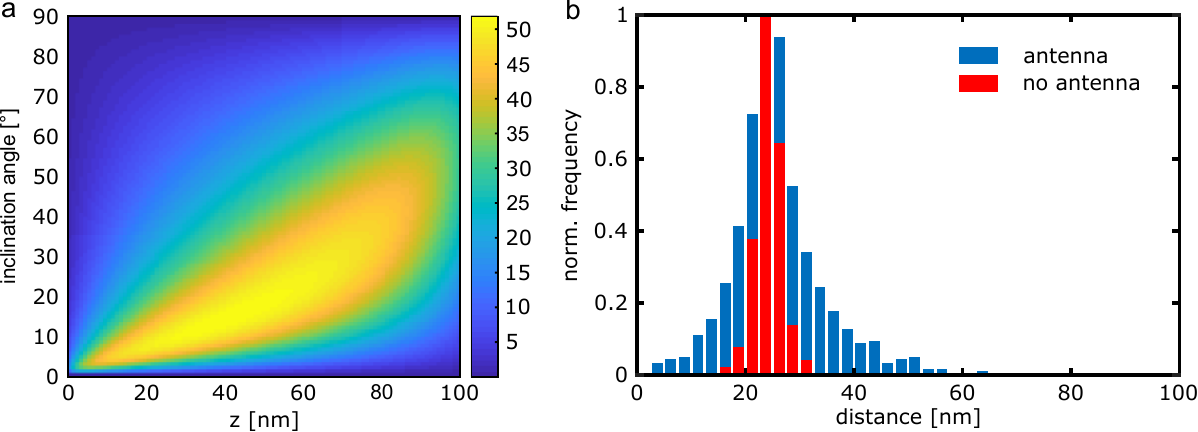}
\caption{\label{SuppFig2}Effect of molecular dipole orientation on localization accuracy and nanometer distance measurements. a) Calculated localization bias in nm as a function of inclination angle and distance from the bottom interface in the antenna geometry of Figure\,\ref{SuppFig1}. b) Simulated distance measurement with and without antenna. The presence of the reflective gold layer and the resulting change of individual PSFs causes a broadening of the distribution.}
\end{figure}

Although the cryogenic use of microscope objectives with high numerical aperture has been demonstrated\cite{Nahmani2017,Furubayashi2019,Wang2019}, their operation imposes some restrictions and, thus, most cryogenic microscopes still use air objectives. This restricts the available signal-to-noise ratio as a consequence of having to image through vacuum. Since the
emission at an interface occurs predominantly into the medium of higher
refractive index the collection efficiency is generally low. For a dipole close
to a cover glass/air interface only $14\%$ of the radiated power falls within
the solid angle of a 0.9 NA objective. In order to mitigate these losses we employ a simple antenna design.
Specifically, we apply a $\SI{100}{nm}$ gold layer on top of a silicon substrate to reflect emitted fluorescence towards the objective. In order to avoid quenching we also apply an $\SI{80}{nm}$ aluminum oxide spacer. The presence of this mirror leads to a threefold enhancement of both the excitation intensity and the power radiated into the upper half-space (see Figure\,\ref{SuppFig1}). However, it also causes a modification of the far-field emission pattern. To investigate the consequences of this effect, we performed simulations for dipoles placed at different distances. Figure\,\ref{SuppFig2}a displays the systematic localization error for a single molecule of arbitrary orientation angle
with respect to the substrate plane and its distance to the mirror. Figure\,\ref{SuppFig2}b shows that co-localization of two emitters at a nominal location of $\SI{23}{nm}$ leads to a broadening of the distribution of separations if the two emitters have random azimuthal orientations. We also found more frequent outliers at smaller and larger distances.

\section*{Sample preparation}

Fast measurements of the photo-physics were performed with single dye molecules embedded in a poly-vinyl alcohol matrix. The dyes were dissolved to nM concentration in a 90:10 mixture of Tris-EDTA buffer with 20 mM magnesium chloride and poly-vinyl alcohol ($5\%$ w/v). Then, $\SI{6}{\mu l}$ of this solution is spin-coated for $\SI{60}{s}$ at $\SI{2000}{rpm}$ onto a glass or gold-coated cover slip. The sample is immediately transferred into the cryostat and cooled down.
For super-resolution imaging a nM stock solution of DNA origamis is diluted 1:90:10 in the same buffer/PVA mixture and spin-coated onto a gold-coated cover slip resulting in a density of approximately $\SI{0.1}{\mu m^2}$. The $\SI{6.5}{nm}$ nanoruler is based on the 12 helix bundle structure\cite{Schmied2014} whereas the $\SI{22.8}{nm}$ ruler is based on a box-shaped design. In both cases we took care to keep origamis in the right buffer conditions to avoid bending or unfolding.     

\section*{Data acquisition} 

Before the start of a new measurement sequence we let the experiment settle for 2h after cool-down to reduce sample drift. We then take 100000 frames at $\SI{70}{Hz}$ for every field-of-view (FOV). The excitation laser intensity was set to $\SI{1}{kW/cm^2}$ to generate the desired low on-off ratio as described in Figure\,\ref{main-Figure1}. 
For polarization modulation measurements the excitation intensity was reduced to $\SI{0.05}{kW/cm^2}$ to avoid saturation and transitions to long-lived dark states. The input polarization is rotated in steps of $2^\circ$ by a motorized linear polarizer with 3 frames taken for $\SI{3}{s}$ at each position. We typically image 3 complete cycles of the modulation to ensure a robust fit to our model.

\section*{Image analysis}  

\begin{figure}[t]
\centering
\includegraphics[width=\textwidth]{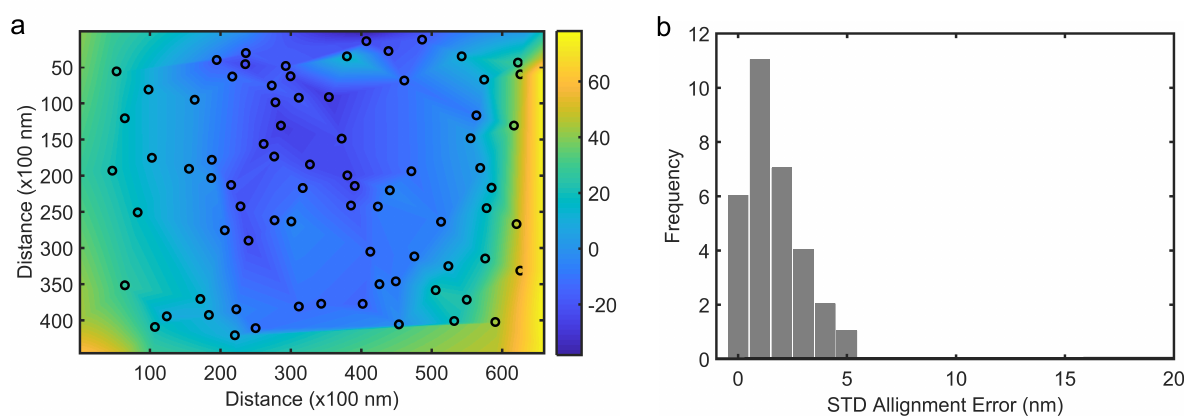}
\caption{\label{SuppFig3}Determining the image registration error. a) Example image of the shift matrix between the two camera images for each polarization channel in the x-direction. The black dots represent the position of the fiducial markers and the color code represents the amount of shift in the x-direction to align both cameras to each other. b) Display of the alignment error of the two cameras.}  
\end{figure}

A 2D median filter with a kernel size of $\SI{5}{\mu m}$ is applied to suppress background fluorescence and PSFs are identified by a threshold-based peak detection algorithm. These PSFs are then localized with a maximum likelihood fit of a conventional 2D Gaussian weighted Poisson noise estimator in each frame on both cameras.
Drift correction is performed by using the single fluorescent emitters as fiducial markers. For each marker the position is tracked throughout the movie and then averaged together with all other markers to perform a drift correction to better than $\SI{1}{nm}$ in the image plane. 
Next, image registration of the polarization channels is performed in two steps. First, a coarse registration aligns all localizations on both cameras to each other. Then, a more precise registration is performed by using the same fiducial markers to generate an alignment matrix. This matrix is established by spatial interpolation of the estimated registration shift of the fiducial markers to correct the remaining error for each position on the cameras. With this procedure we could limit the registration error to a median value of $\SI{1.5}{nm}$. This error was calculated by taking the median of the alignment errors of individual fiducial markers. To this end we perform the correction with a shift map that explicitly excludes one particular marker. By repeating this step for all fiducial markers we obtain the histogram shown in Figure\,\ref{SuppFig3}. 
After drift correction and image registration localizations are assigned to a fluorophore according to the estimated polarization. To this end, we determine the two peak positions in the polarization histogram and pick frames where the value is smaller or larger than the peak value.  
To further limit false assignments due to a small residual overlap we require a minimum separation of the peak values of $15^\circ$. As indicated in Figure\,\ref{main-Figure4} of the main manuscript, we require a localization precision better than $\SI{1}{nm}$ for both fluorophores.
Data from polarization modulation measurements were analyzed in a similar fashion. However, here we only require a coarse image registration on the pixel scale to identify the same PSF in the two polarization channels as well as on the localization camera. For statistical analysis of distances we only kept data points with an uninterrupted trace. 

\section*{Estimation of distance and uncertainty}

\begin{figure}[t]
\centering
\includegraphics[width=\textwidth]{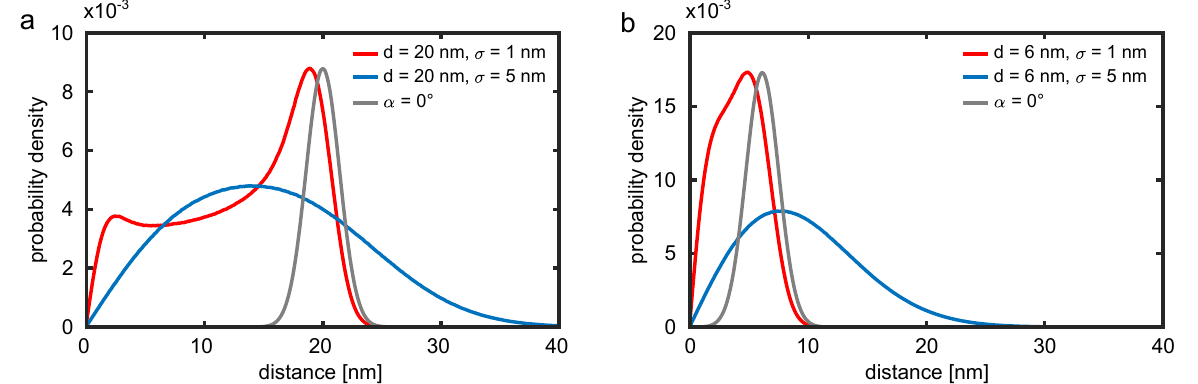}
\caption{\label{SuppFig4}Extended model of the distance distribution between two fluorophores. The Rician distribution describes the effect of finite localization uncertainty while its convolution with a $\cos\alpha$ function also accounts for arbitrary 3D orientation. Tilting of a nanoruler with respect to the optical axis leads to a shorter apparent distance between the fluorophores in the image plane. a) For a true distance of 
$\SI{20}{nm}$ we compare the expected distributions for arbitrary 3D orientation at two different localization uncertainties (red and blue) with only flat-lying nanorulers (gray line). As the experimental distributions in Figure\,\ref{main-Figure4} do not show the features of arbitrary orientation we infer that they are mostly parallel to the surface. b) The same comparison for a true distance of $\SI{6}{nm}$.}
\end{figure}

In order to estimate the distance between the localized sites in an ensemble of particles one has to keep several considerations in mind. First, our technique relies on averaging many particles and therefore assumes a homogeneous ensemble. Second, the distance distribution is generally asymmetric\cite{Churchman2006}. Third, depending on the sample preparation one might have to consider the 3D orientation of the particles. In Figure\,\ref{SuppFig4} we show the theoretical distance distributions of a model that allows for inclination of a nanoruler outside the image plane\cite{Weisenburger2017}. One can see clearly that a $\SI{20}{nm}$ ruler with arbitrary 3D orientation shows distributions distinct from our experimental data. For completeness, we also show the distributions of a $\SI{6}{nm}$ ruler.    
In the uncertainty of the distance measurements we also considered several contributions. First, we have to take the calibration of the pixelsize of our cameras with an uncertainty of $\SI{1}{nm}$ into account. Next, we included an estimation error of the model parameter $\sigma$ on the fitted distance as shown in Figure\,\ref{main-Figure4}e. Here, the largest contribution to the uncertainty is due to the dipole error. Furthermore, the influence of false positive detection events through occasional failures of the analysis routine was excluded by performing manual particle picking.
These statistically independent contributions add to give at a final distance uncertainty of about $\SI{2}{nm}$ in our experiment.
Lastly, it should be pointed out that DNA origamis can exhibit significant structural heterogeneity leading to an additional uncertainty contribution that is difficult to characterize but may depend on the buffer environment\cite{Schmied2014} and even exceed $\SI{1}{nm}$.

\bibliography{references}